\documentclass{aastex}

\usepackage{emulateapj5,mathptmx}

\makeatletter
\newenvironment{inlinetable}{%
\def\@captype{table}%
\noindent\begin{minipage}{0.999\linewidth}\begin{center}\footnotesize}
{\end{center}\end{minipage}\smallskip}

\newenvironment{inlinefigure}{%
\def\@captype{figure}%
\noindent\begin{minipage}{0.999\linewidth}\begin{center}}
{\end{center}\end{minipage}\smallskip}
\makeatother

\def\M324{\ensuremath{M_{g,324}}}

\begin{document}

\lefthead{EVOLUTION OF CLUSTER SCALING RELATIONS}
\righthead{VIKHLININ ET AL.}

\title{Evolution of the cluster X-ray scaling relations since $z>0.4$}

\author{A. Vikhlinin\altaffilmark{1}\altaffilmark{,2}, L.\ 
  VanSpeybroeck\altaffilmark{1}, M.  Markevitch\altaffilmark{1},
  W.~R.~Forman\altaffilmark{1}, L. Grego\altaffilmark{1}}

\altaffiltext{1}{Harvard-Smithsonian Center for Astrophysics, 60 Garden St.,
  Cambridge, MA 02138; avikhlinin@cfa.harvard.edu} \altaffiltext{2}{Also Space
  Research Institute, Moscow, Russia}

\begin{abstract}
  
  We derive correlations between X-ray temperature, luminosity, and gas mass
  for a sample of 22 distant, $z>0.4$, galaxy clusters observed with
  Chandra. We detect evolution in all three correlations between $z>0.4$ and
  the present epoch. In particular, in the $\Omega=0.3$, $\Lambda=0.7$
  cosmology, the luminosity corresponding to a fixed temperature scales
  approximately as $(1+z)^{1.5\pm0.3}$; the gas mass for a fixed luminosity
  scales as $(1+z)^{-1.8\pm0.4}$; and the gas mass for a fixed temperature
  scales as $(1+z)^{-0.5\pm0.4}$ (all uncertainties are 90\% confidence). We
  briefly discuss the implication of these results for cluster evolution
  models.
\end{abstract}


\section{Introduction}

Correlations between the X-ray properties of galaxy clusters are useful
statistical tools which allow to study the cluster physics. The best-studied
correlation for the low-redshift clusters is that of the X-ray luminosity
with the temperature (e.g., Mushotzky 1984, David et al.\ 
1993, Markevitch 1998, Arnaud \& Evrard 1999).  The mass of the intracluster
gas correlates with both temperature (Mohr, Mathiesen \& Evrard 1999,
Vikhlinin, Forman \& Jones 1999) and X-ray luminosity (Voevodkin, Vikhlinin
\& Pavlinsky 2002b). The observed tightness of these correlations indicates
a similar formation history for all clusters and is consistent with the
predictions of the self-similar models of the cluster formation.

However, the details of these scaling relations are different from the
predictions of the self-similar theory. The best-known example is the slope
of the $L-T$ correlation: it is observed that $L\propto T^{2.7}$ for hot
clusters (e.g., Markevitch 1998), while theory predicts $L\propto T^{2}$
(e.g., Kaiser 1981).  Such deviations may be due to non-gravitational
processes such as preheating (e.g., Cavaliere, Menci \& Tozzi 1997) or
radiative cooling and feedback from star formation (Voit \& Bryan 2001).
Observing the evolution of scaling relations can provide useful constraints
on such models. The scaling relations at high redshift also are of great
value for cosmological studies based on cluster evolution. They provide the
means to convert an easily observed X-ray luminosity function into the more
cosmologically useful temperature or mass functions (e.g., Borgani et al.\ 
2001).


Most of the previous high redshift studies have focused on the $L-T$
relation. Mushotzky \& Scharf (1997) analyzed a large sample of distant
clusters observed with ASCA (most at $z\sim 0.3$, with a few at $z>0.4$) and
found no evidence for evolution in the $L-T$ relation. Several results for a
small number of distant cluster observed by \emph{Chandra} have been
published recently. Borgani et al.\ (2001) analyzed a sample of 7 clusters
at $z>0.5$ and concluded that the data allow at most a very mild evolution
--- if the luminosity for the given temperature is $L(z)\propto(1+z)^A$,
then $A<1$. A similar conclusion has been reached by Holden et al. (2002)
from an analysis of 12 clusters at $z>0.7$.

The $L-T$ relation at both low and high redshifts has a large intrinsic
scatter, comparable to the expected evolutionary effects. The scatter in the
low-redshift $L-T$ relation is significantly reduced when the central
cooling regions of the clusters are excised from both the luminosity and
temperature measurements (Fabian et al.\ 1994, Markevitch 1998). Therefore,
it is desirable to exclude the cooling cores in the distant clusters because
this too may reduce the scatter and thus more easily expose any evolution.
This task is feasible only with the \emph{Chandra}'s arcsecond angular
resolution.

As of Spring 2002, \emph{Chandra} had observed 22 clusters at $z>0.4$ with
sufficient exposure for accurate temperature measurements. Most of this
sample is derived from flux-limited X-ray surveys: 7 clusters from the EMSS
(Henry et al.\ 1992), 11 from the \emph{ROSAT} serendipitous surveys, 160
deg$^2$ (Vikhlinin et al.\ 1998), RDCS (Rosati et al.\ 1998), WARPS (Ebeling
et al. 2000), and 2 from the \emph{ROSAT} All-Sky Survey. We use these
\emph{Chandra} observations to derive correlations between the X-ray
luminosity, temperature, and gas mass at $z>0.4$.

We use $H_0=50$~km~s$^{-1}$~Mpc$^{-1}$ throughout.

\begin{table*}
\def\j{\phantom{1}}
\caption{Cluster sample}
{\centering
\footnotesize

\begin{tabular}{lccrrc}
\hline
\hline
Name            & $z$   & $T$  & $L_{0.5-2}^a$ & \multicolumn{1}{c}{$L_{\rm bol}^b$}  &
$\M324$ \\ 
                &       & (keV)&  &  &  $(10^{14}\, M_\odot)$ \\

\hline
MS 0016+1609     & 0.541 &  $\j 9.9\pm0.5$ &  22.8 &  113.3&  $ 6.43\pm 0.65$  \\
MS 0302+1658     & 0.424 &  $\j 3.6\pm0.5$ &   4.7 &   10.6&  $ 1.07\pm 0.40$  \\
MS 0451--0305    & 0.537 &  $\j 8.1\pm0.8$ &  20.7 &   91.7&  $ 3.68\pm 0.77$  \\
MS 1054--0321    & 0.823 &  $\j 7.8\pm0.6$ &  16.5 &   70.9&  $ 2.58\pm 0.37$  \\
MS 1137+6625     & 0.782 &  $\j 6.3\pm0.4$ &   8.4 &   32.4&  $ 1.41\pm 0.28$  \\
MS 1621+2640     & 0.426 &  $\j 7.6\pm0.9$ &   6.3 &   27.0&  $ 2.89\pm 0.62$  \\
MS 2053-0449     & 0.583 &  $\j 5.2\pm0.7$ &   3.5 &   10.8&  $ 0.95\pm 0.32$  \\
CL 1120+2326     & 0.562 &  $\j 4.8\pm0.5$ &   3.7 &   12.5&  $ 1.19\pm 0.27$  \\
CL 1221+4918     & 0.700 &  $\j 7.2\pm0.6$ &   7.0 &   28.7&  $ 2.01\pm 0.36$  \\
CL 1416+4446     & 0.400 &  $\j 3.7\pm0.3$ &   4.2 &    8.9&  $ 1.42\pm 0.38$  \\
CL 1524+0957     & 0.516 &  $\j 5.1\pm0.6$ &   4.5 &   15.7&  $ 1.67\pm 0.40$  \\
CL 1701+6421     & 0.453 &  $\j 5.8\pm0.5$ &   4.9 &   15.9&  $ 1.81\pm 0.47$  \\
CL 0848+4456     & 0.574 &  $\j 2.7\pm0.3$ &  10.6 &   38.8&  $ 0.36\pm 0.13$  \\
WARPS 0152--1357 & 0.833 &  $\j 5.8\pm0.6$ &   1.2 &    3.1&  $ 2.88\pm 0.55$  \\
RDCS 0848+4452   & 1.261 &  $\j 4.7\pm1.0$ &   1.8 &    6.0&  $ 0.20\pm 0.08$  \\
RDCS 0910+5422   & 1.100 &  $\j 3.5\pm0.7$ &   2.0 &    5.9&  $ 0.26\pm 0.11$  \\
RDCS 1317+2911   & 0.805 &  $\j 2.2\pm0.5$ &   0.8 &    2.0&  $ 0.21\pm 0.09$  \\
RDCS 1350.0+6007 & 0.805 &  $\j 4.3\pm0.6$ &   4.2 &   13.2&  $ 1.04\pm 0.33$  \\
RASS 1347--114   & 0.451 &  $  14.1\pm0.9$ &  60.1 &  260.4&  $ 8.77\pm 1.60$  \\
RASS 1716+6708   & 0.813 &  $\j 6.6\pm0.8$ &   7.2 &   28.8&  $ 1.25\pm 0.33$  \\
3C295            & 0.460 &  $\j 5.3\pm0.5$ &   9.1 &   16.3&  $ 1.51\pm 0.48$  \\
CL0024+17        & 0.394 &  $\j 4.8\pm0.6$ &   3.1 &    9.2&  $ 1.24\pm 0.37$  \\
\hline
\end{tabular}
\par

\medskip

\begin{minipage}{0.8\linewidth}
\footnotesize

$^a$ --- Total X-ray luminosity in the 0.5--2 keV band within the 2~Mpc
radius, $10^{44}\,$erg~s$^{-1}$.  $^b$ --- Bolometric luminosity within the
2~Mpc radius excluding the central cooling regions, $10^{44}\,$erg~s$^{-1}$.
All quantities are computed for the $\Omega=0.3$, $\Lambda=0.7$ cosmology.

\end{minipage}
\par
}

\end{table*}

\section{Data Analysis}

\emph{Chandra} data were reduced in a standard manner (see Markevitch \&
Vikhlinin 2001 for a fuller description of our procedures). Spectral
analysis was performed in the 0.8--10~keV band, to minimize the calibration
uncertainties due to time-dependent decrease in the low energy quantum
efficiency\footnote{cxc.harvard.edu $\rightarrow$ Calibration $\rightarrow$
  ACIS $\rightarrow$ April HEAD Meeting}. The effective area below 1.8 keV
in the front-illuminated CCDs was corrected by an empirical factor of 0.93
to improve the cross-calibration with the back-illuminated CCDs (see
Markevitch \& Vikhlinin 2001 for details).  Background maps were generated
using the prescriptions of Markevitch (2002). Cluster spectra were fit with
the MEKAL model, with absorption fixed at the Galactic value, and abundance
fixed at $0.3$ Solar unless the quality of the data allowed a direct
measurement.

The imaging analysis was performed in the 0.7--2~keV energy band, but we
also checked 2--7\,keV images to excise hard, self-absorbed sources which
may contaminate the spectral analysis of the faintest clusters.

Our aim was to compare the high-redshift $L-T$ relation with the
low-redshift one from Markevitch (1998), and so we followed this analysis as
closely as possible. Temperatures were measured by fitting a spectrum
integrated within a radius of 0.5--1~Mpc with a single-temperature MEKAL
model; this region contained at least 75\% of the total cluster flux. The
bolometric X-ray luminosity was computed by normalizing the best-fit
spectral model by the observed total 0.7--2~keV count rate within 2~Mpc.
The extrapolation of a typical $\beta$-model profile shows that at most 4\%
of the X-ray flux originates outside this radius, which we ignore. The
clusters with sharply peaked surface brightness profiles are likely to
contain cooling cores, and so we excluded the central 100~kpc regions in
such clusters from both the spectral analysis and count rate computations,
multiplying the measured count rates by 1.06 to account for the flux within
$r<100$~kpc in a typical $\beta$-model cluster.

\begin{inlinefigure}
  \includegraphics[width=0.85\linewidth]{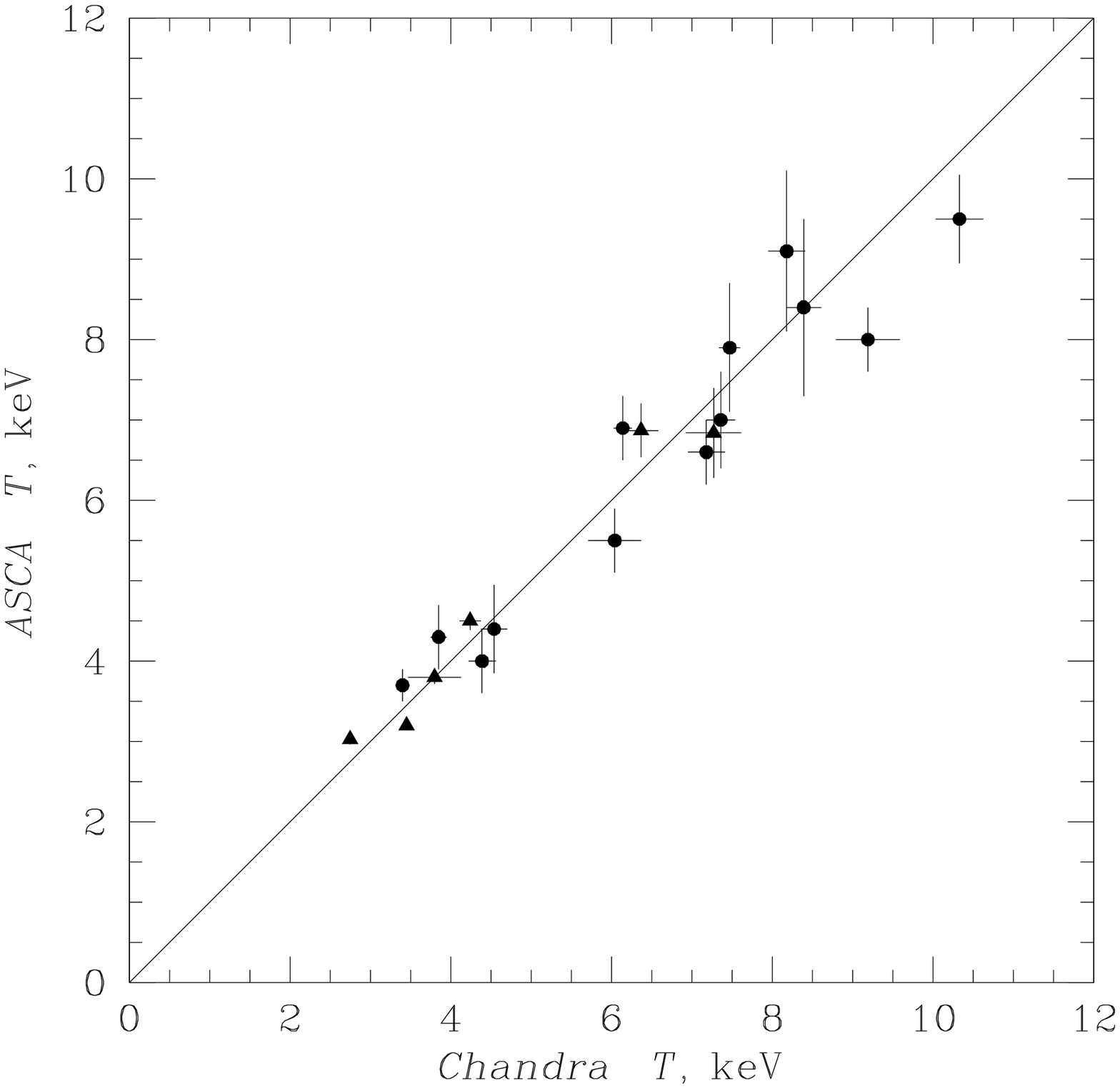}
  \caption{Comparison of \emph{Chandra} and \emph{ASCA}
    temperatures. \emph{ASCA} cooling flow corrected values for A401, A478,
    A644, A754, A85, A780, A2029, A2256, A2597, A3376, A3558, A3667, and
    MKW3s are from Markevitch (1998; circles). Triangles represent non-
    or weak cooling flow clusters A1060, A2147, A2218, A2255, and AWM7 with
    the \emph{ASCA} values from White (2000) without any cooling flow
    correction. Uncertainties are 90\% confidence.  }
\label{fig:tcal}
\end{inlinefigure}

It is important to verify that there is no systematic difference between our
\emph{Chandra} and earlier \emph{ASCA} temperature measurements. Using the
procedure outlined above, we derived temperatures for a number of nearby
clusters from the \emph{ASCA} samples of Markevitch (1998) and White (2000)
and confirmed that on average, the agreement between \emph{Chandra} and
\emph{ASCA} measurements is at a level of 5\% or better
(Fig.~\ref{fig:tcal}).

\begin{figure*}[htb]
\centerline{
\includegraphics[width=0.45\linewidth]{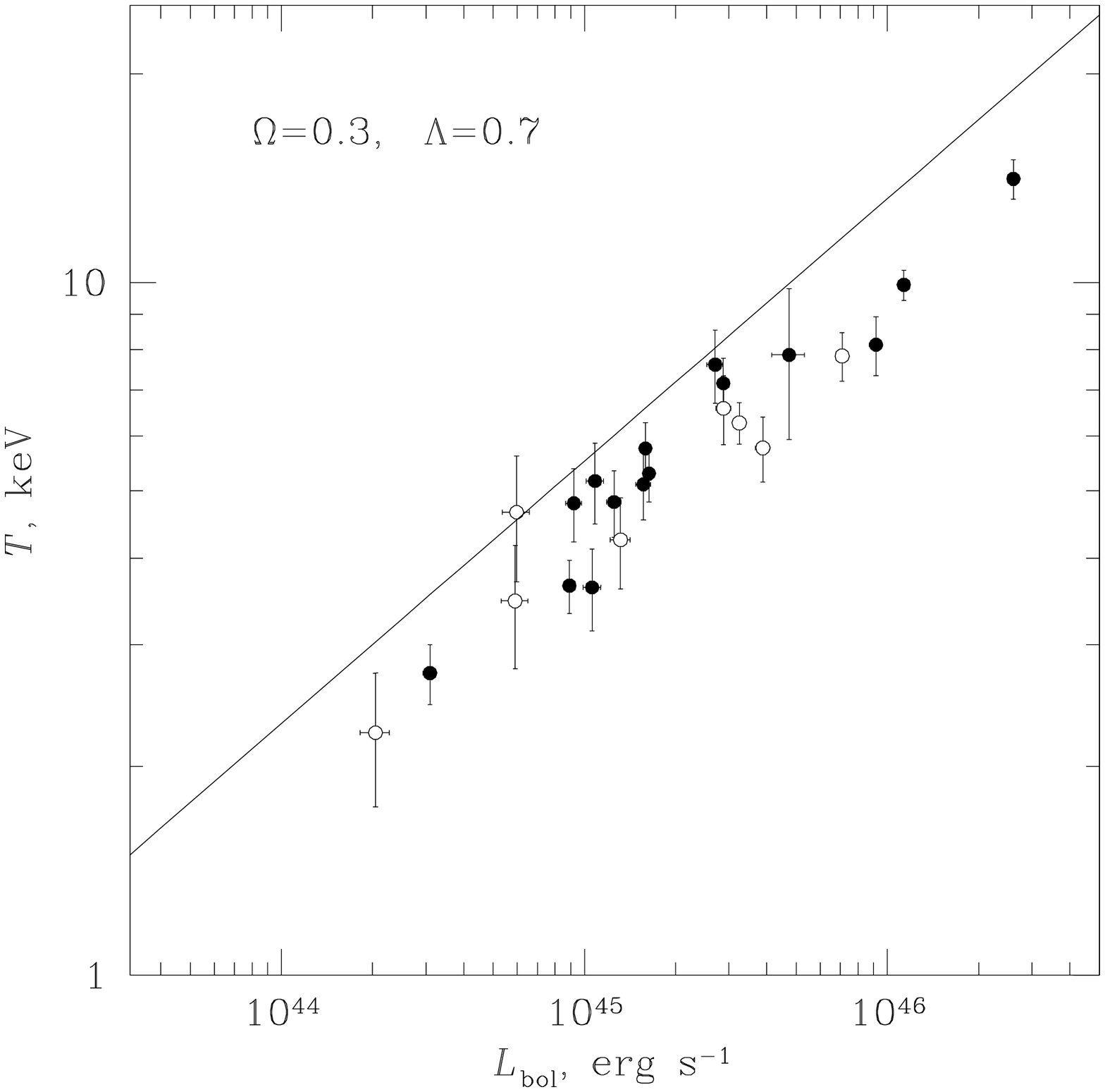}
\hfill
\includegraphics[width=0.45\linewidth]{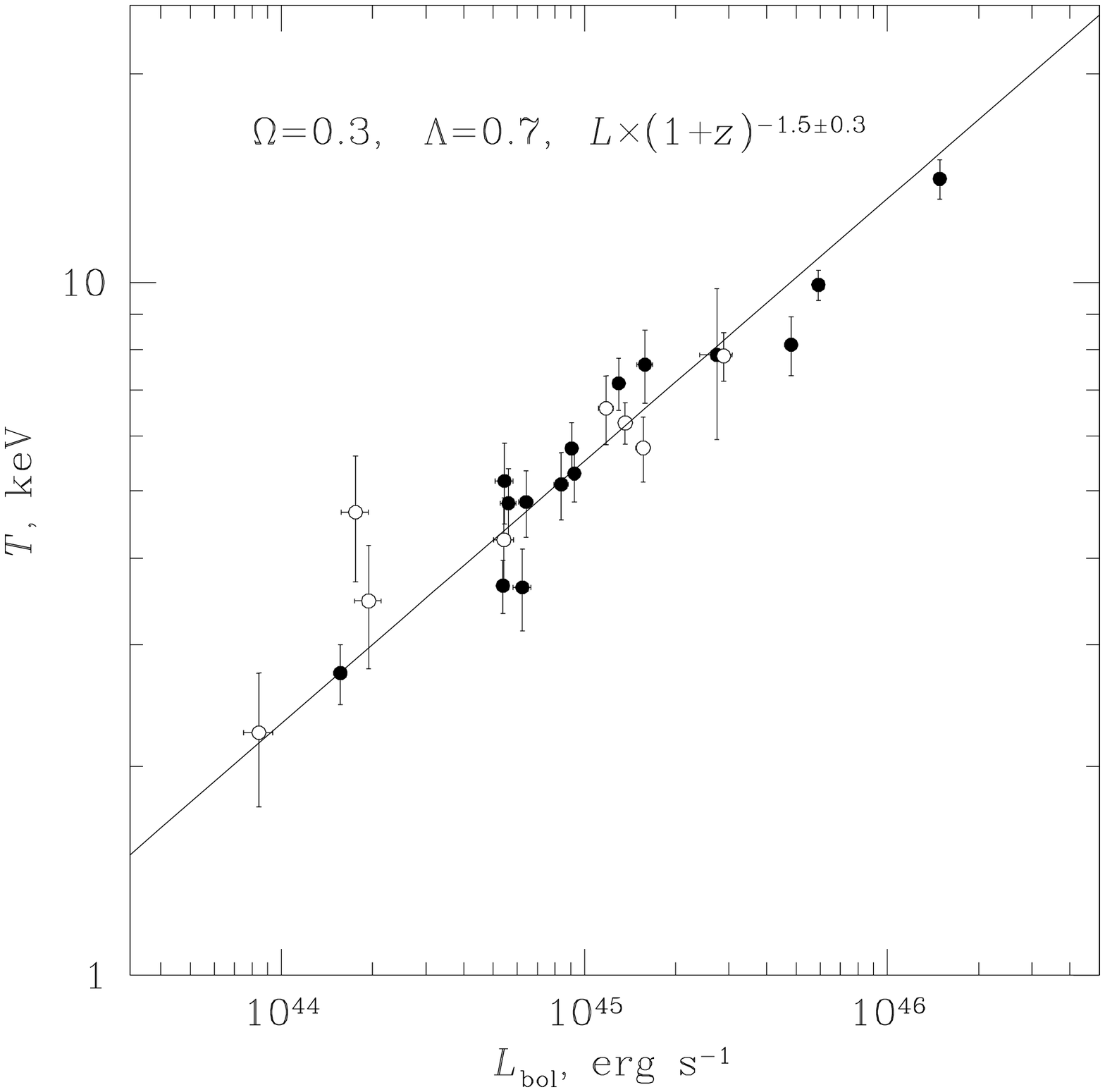}
}
\caption{Correlation of the cluster temperature and bolometric
  luminosity. Both quantities are measured excluding any central cooling
  regions of the clusters. Filled circles correspond to clusters with
  $0.4<z<0.7$, and open circles to clusters at $z>0.7$. The solid line shows
  the $L-T$ correlation for the low-redshift clusters (Markevitch 1998). 
  The right panel shows the effect of removing the best-fit evolution given
  in Table~\ref{tab:pars}.
}
\label{fig:lt}
\end{figure*}

\section{Results}

The resulting $L-T$ correlation in the $\Omega=0.3$, $\Lambda=0.7$ cosmology
is shown in Fig.~\ref{fig:lt}. The scatter in the correlation is very small
--- it is almost consistent with the measurement uncertainties. The slope is
consistent with the low-redshift relation, but the normalization is
significantly different --- for the same temperature, clusters at high
redshift are more luminous. If the $z$-dependence of the correlation is
parameterized as $L \propto (1+z)^{A_{LT}}\, T^\alpha$ with $\alpha$ fixed
at its local value of 2.64, then $A_{LT} = 1.5\pm0.3$ (90\% confidence) for
this cosmology. For $\Omega=1$, $\Lambda=0$, evolution of the $L-T$ relation
is weaker, but still significant (Table \ref{tab:pars}).

Let us now consider correlations involving the gas mass. Assuming spherical
symmetry, the radial profile of the gas mass in a cluster is trivially
derived from the observed X-ray surface brightness profile. The main
question is: within which radius to measure the gas mass? We chose the
radius defined by the mean gas overdensity over the average baryon density
in the Universe $\delta_g = M_g (r) / \left(4/3\,\pi\, r^3
  \langle\rho_b\rangle\right) = 324\, (h/0.5)^{1/2}$, where $M_g (r)$ is the
gas mass within the radius $r$, and $\langle\rho_b\rangle =
5.55\,M_\odot~\mbox{kpc}^{-3} \times (1+z)^3$ is the average baryon density
from Big Bang nucleosynthesis theory (Burles, Nollett, \& Turner 2001). This
choice is motivated by two considerations. First, if the mass fraction of
baryons in clusters is representative of that for the Universe as a whole,
then at large distances from the cluster center, $\delta_g$ should equal the
mean total mass overdensity, $\delta_m$, and at the same time $\delta_g$ is
easily determined from the X-ray imaging data (our procedure is described in
Vikhlinin et al.\ 1999). Strictly speaking, the baryon mass should include
the stellar material but we ignore this minor contributor.  Second, Jenkins
et al.\ (2001) developed a universal model for the cluster total mass
function for masses corresponding to $\delta_m = 324$ at the redshift of
observation, therefore our values can be readily compared to theoretical
predictions.

\begin{inlinefigure}
\centerline{
\includegraphics[width=0.85\linewidth]{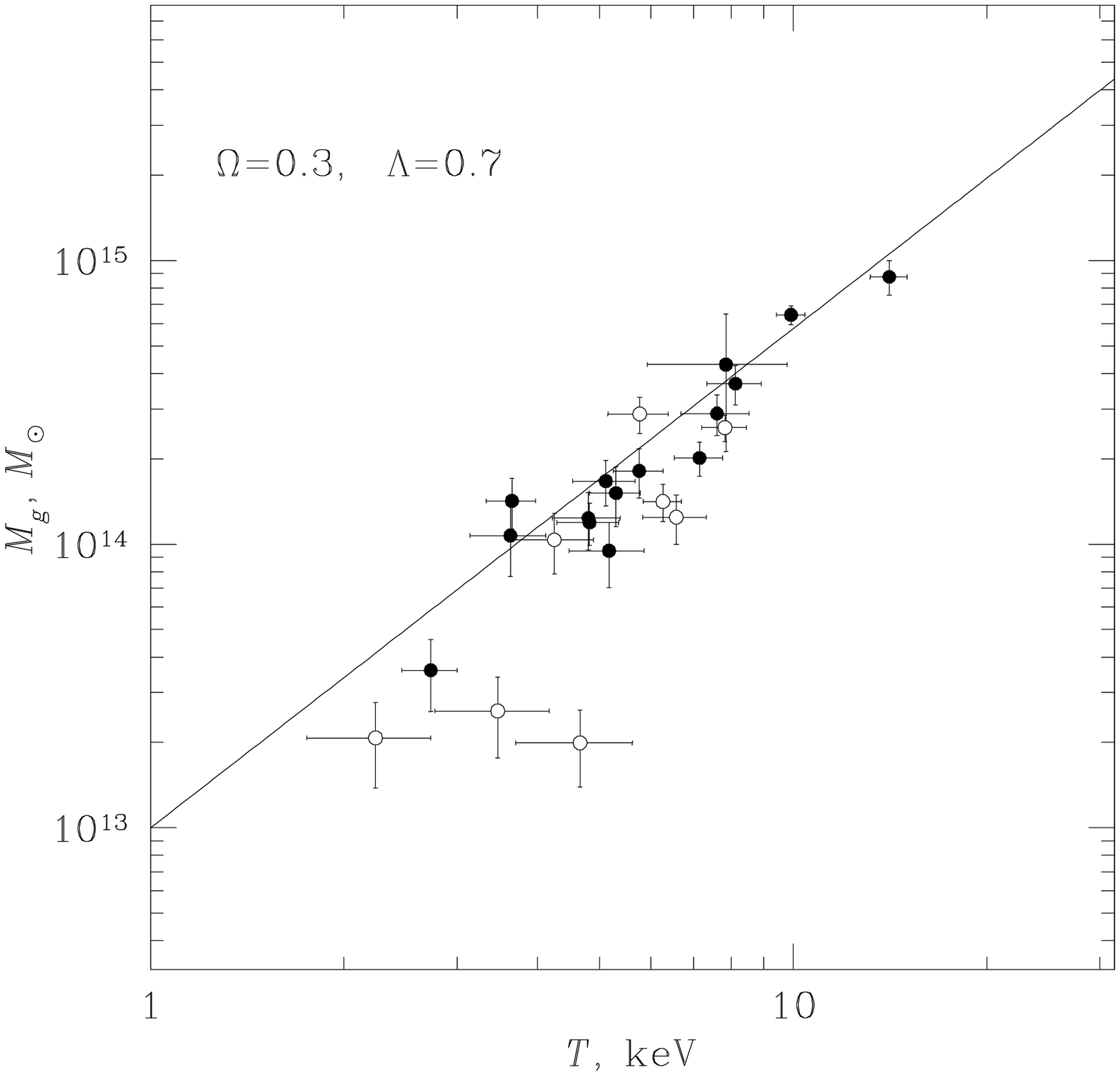}
}
\caption{Correlation of cluster temperature and gas mass within the
  radius of mean overdensity of 324. Symbols same as in Fig.~\ref{fig:lt}.
  Solid line shows the $M_{\rm gas}-T$ correlation for the low-redshift
  clusters (Voevodkin et al.\ 2002a). }
  \label{fig:mt}
\end{inlinefigure}

Figure~\ref{fig:mt} shows the correlation of \M324\ with the cluster
temperature. The solid line shows this relation for a sample of low-redshift
clusters (Voevodkin et al.\ 2002a). For the $\Omega=0.3$, $\Lambda=0.7$
cosmology, the evolution is very weak. If the $z$-dependence in the
correlation is parameterized as $M_g \propto (1+z)^{A_{MT}}\, T^\beta$ with
$\beta$ fixed on its local value of 1.76, then $A_{MT} = 0.5\pm0.4$. For
$\Omega=1$, $\Lambda=0$, the derived gas masses are significantly smaller
and there is a strong evolution in the $M_g-T$ relation, $A_{MT}=1.5\pm0.4$.

\begin{figure*}[htb]
\centerline{
\includegraphics[width=0.45\linewidth]{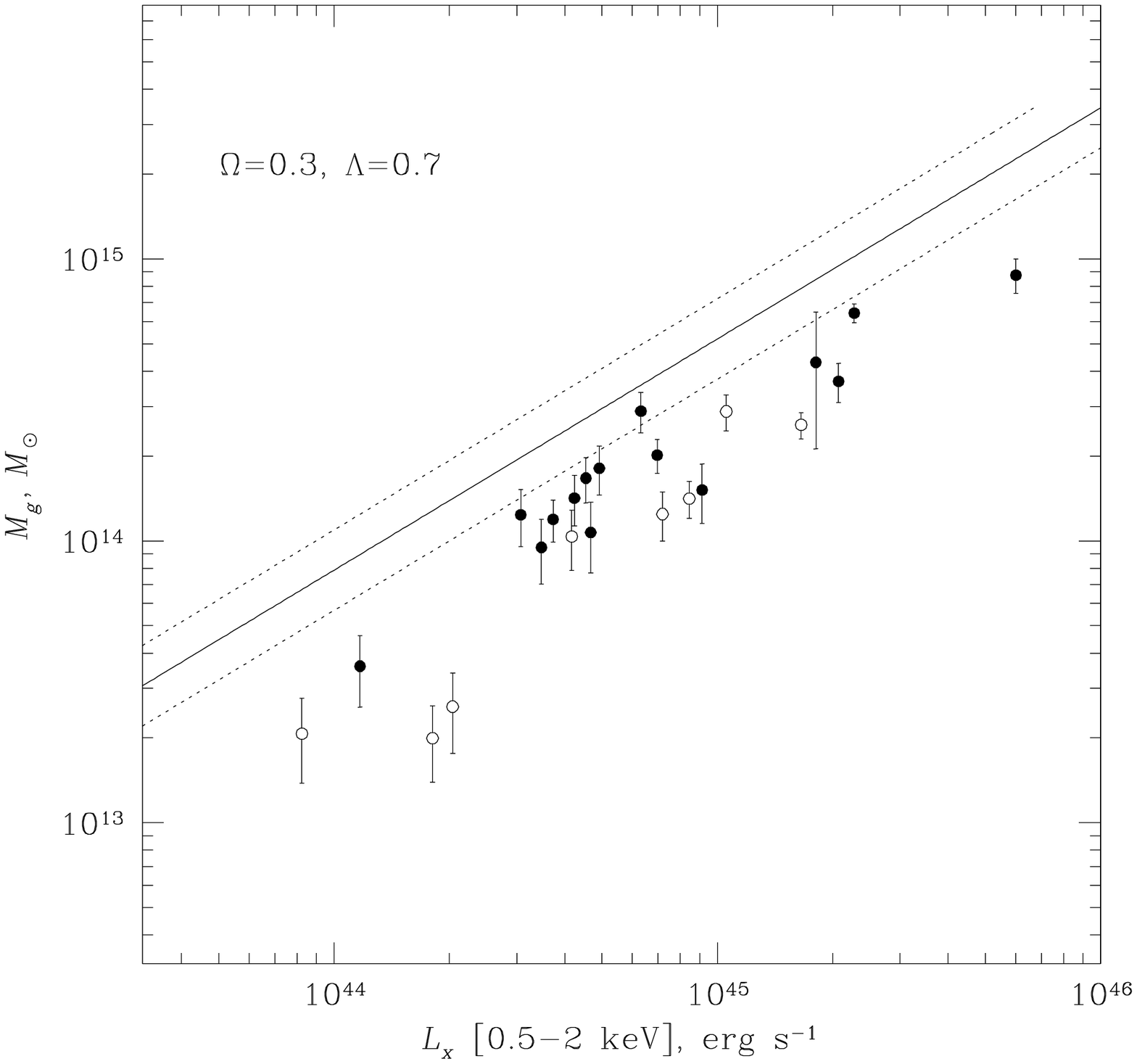}
\hfill
\includegraphics[width=0.45\linewidth]{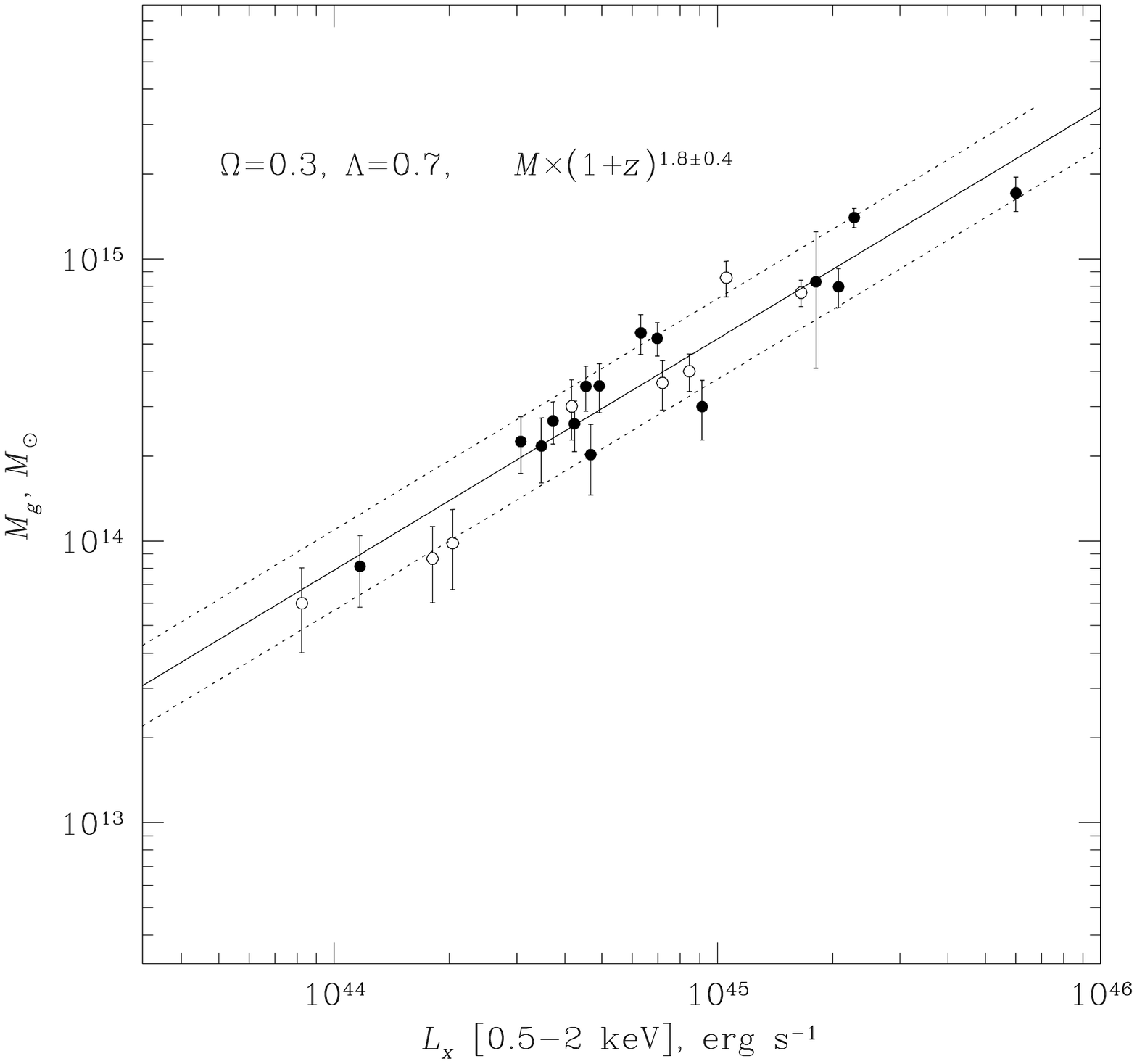}
}
\caption{Correlation of cluster luminosity in the 0.5--2 keV band and
  gas mass within the radius of mean overdensity of 324. Symbols same as in
  Fig.~\ref{fig:lt}.  The luminosities \emph{include} the central cooling
  regions. Solid line shows the $M-L$ correlation for low-redshift clusters,
  and dotted lines correspond to a 1-$\sigma$ scatter about the mean
  relation (Voevodkin et al.\ 2002b). The right panel shows the effect of
  removing the best-fit evolution given in
  Table~\ref{tab:pars}.}\label{fig:ml}
\end{figure*}

Figure~\ref{fig:ml} shows a correlation of \M324\ with the X-ray luminosity.
This relation is potentially useful for estimating the mass function from
the X-ray luminosity function for distant clusters. Therefore, we used in
this case the total cluster luminosity (i.e. not excluding the central
regions) in the 0.5--2~keV energy band --- the quantity commonly measured in
the serendipitous \emph{ROSAT} cluster surveys. The solid line in
Fig.~\ref{fig:ml} shows the $M-L_{\rm tot}$ correlation for low-redshift
clusters (Voevodkin et al.\ 2002b). The $M_g-L$ relation shows a strong evolution
in all cosmologies.  If its $z$-dependence is parameterized as a power law,
$M\propto (1+z)^{A_{ML}}\,L^\gamma$ with $\gamma$ fixed at its local value
of $0.83$, then $A_{ML}\approx 2\pm0.4$ (Table~\ref{tab:pars}).

\begin{inlinetable}
\caption{Evolution of scaling relations in different cosmologies}\label{tab:pars}
\footnotesize

\def\arraystretch{1.5}

{\centering
\begin{tabular}{lcccc}
\hline
\hline
Cosmology & $q_0$ & $A_{LT}$ & $A_{MT}$ & $A_{ML}$ \\
\hline
$\Omega=0.3$, $\Lambda=0.7$ & $-0.55$                     & $1.5\pm0.3$ & $0.5\pm0.4$ & $1.8\pm0.4$\\
$\Omega=1$, $\Lambda=0$     & $\phantom{-}0.5\phantom{5}$ & $0.6\pm0.3$ & $1.5\pm0.4$ & $2.1\pm0.4$\\
\hline
\end{tabular}
\par
}
\medskip
\begin{minipage}{0.99\linewidth}
For other cosmologies, the power law slopes in the relations $L\propto
(1+z)^{A_{LT}}\,T^\alpha$, $M_g\propto (1+z)^{A_{MT}}\,T^\beta$, $M_g\propto
(1+z)^{A_{ML}}\,L^\gamma$ can be approximated by interpolating over $q_0$.
Uncertainties are at the 90\% confidence level for a single parameter.
\end{minipage}
\end{inlinetable}

\section{Discussion and conclusions}

We have used \emph{Chandra} observations of a sample of 22 distant, $z>0.4$,
clusters to show that the correlations between the cluster temperature,
luminosity, and gas mass evolve significantly with respect to the
low-redshift relations.

Our detection of significant evolution in the $L-T$ relation appears to
contradict some other recent \emph{Chandra} studies (Borgani et al.\ 2001,
Holden et al.\ 2002). The difference between their and our results can be
traced mostly to a more consistent comparison of the high- and low-redshift
samples, such as exclusion of the cool cores and extraction of the
luminosities in the 2~Mpc aperture. Note that the results of Novicki et al.\ 
(2002) who self-consistently use \emph{ASCA} data for the nearby and distant
clusters, agree well with our $L-T$ evolution.

It is theoretically expected that within clusters, the baryon contribution
to the total mass, $f_b$, should be close to the average value in the
Universe (e.g., White et al.\ 1993).  This notion continues to gain
observational support (most recently, Allen, Schmidt \& Fabian 2002). If
$f_b$ is indeed constant, the evolutions in the $M-L$ and $M-T$ relations
involving the gas mass or total mass should be identical.

The observed evolution of the cluster $M_{g}-T-L$ correlations indicates
that clusters at high redshift were systematically denser than at present
--- hotter and more luminous for a given mass, as expected in a theory of
the hierarchical self-similar formation. However, the details of the
observed evolution contradict the self-similar predictions. For example, the
standard theory (e.g., Bryan \& Norman 1998) predicts that for a given
temperature, the product $H(z) M_\Delta\,(T,z)$ should be constant, where
$M_\Delta$ is the mass measured within the radius of the overdensity
$\Delta$ with respect to the critical density at redshift $z$, and $H(z)$ is
the Hubble constant. For the realistic cluster density profiles, this
implies that approximately $M_\delta(z,T) \propto (1+z)^{-3/2}$ in almost
any cosmology, where $M_\delta$ corresponds to the mean overdensity $\delta$
relative to the average density at redshift $z$. However, we observe little
evolution in the $M_{g,\delta}-T$ relation for the currently favored
$\Omega=0.3$, $\Lambda=0.7$ cosmology, seemingly at odds with the
theoretical prediction. This possibly indicates the importance of
non-gravitational processes for heating the intracluster gas that would
change $T$ but are unlikely to modify $M_g$, just as is inferred from the
scaling relations for low-redshift clusters.

\acknowledgements

This work was supported by NASA grant NAG5-9217 and contract NAS8-39073. We
thank J.~P.~Henry and M.~Novicki for sharing their results prior to
publication.

\end{document}